\documentclass[journal abbreviation, manuscript]{copernicus}
\usepackage[dvipsnames]{xcolor}
\usepackage[mathscr]{euscript}
\usepackage{amssymb,amsmath,amsthm,stmaryrd}
\usepackage{graphicx}

\usepackage{natbib}
\allowdisplaybreaks

\newcommand{\argmin}{\mathop{\arg\min}}
\newcommand{\barSWI}{\overline{\text{SWI}}}
\newcommand{\bbR}{\mathbb{R}}
\newcommand{\calA}{\mathcal{A}}

\newcommand{\calE}{\mathcal{E}}

\newcommand{\calG}{\mathcal{G}}
\newcommand{\calI}{\mathcal{I}}
\newcommand{\calL}{\mathcal{L}}
\newcommand{\calO}{\mathcal{O}}
\newcommand{\calX}{\mathcal{X}}
\newcommand{\calZ}{\mathcal{Z}}
\newcommand{\Exp}{\mathbb{E}}

\begin{document}
\nolinenumbers
\title{Forecasting the cost of drought events in France by Super Learning}


\Author[1,2,*]{Geoffrey}{Ecoto}
\Author[2,*]{Antoine}{Chambaz}

\affil[1]{Caisse Centrale de Réassurance}
\affil[2]{Université Paris Cité, CNRS, MAP5, F-75006 Paris, France}
\affil[*]{These authors contributed equally to this work.}




\correspondence{Geoffrey Ecoto (gecoto@ccr.fr)}

\runningtitle{Forecasting the cost of drought events by Super Learning}

\runningauthor{G. Ecoto \& A. Chambaz}

\received{}
\pubdiscuss{} 
\revised{}
\accepted{}
\published{}


\firstpage{1}

\maketitle

\begin{abstract}
  Drought events are the second most expensive type of natural disaster within
  the  French legal  framework  known as  the  natural disasters  compensation
  scheme.   In recent  years, drought  events  have been  remarkable in  their
  geographical scale and intensity.  We develop and apply a new methodology to
  forecast the cost  of a drought event in France.   The methodology hinges on
  Super Learning~\citep{SL2007,Benkeser2018},  a general  aggregation strategy
  to learn a feature  of the law of the data  identified through an \textit{ad
    hoc} risk function by relying on  a library of algorithms.  The algorithms
  either compete  (discrete Super  Learning) or collaborate  (continuous Super
  Learning), with  a cross-validation  scheme determining the  best performing
  algorithm  or combination  of algorithms,  respectively.  Our  Super Learner
  takes into account  the complex dependence structure induced in  the data by
  the spatial and temporal nature of drought events.
\end{abstract}


\introduction  
\label{sec:intro}

In  this  study we  call  \textit{a  drought  event}  the phenomenon  of  clay
shrinking and  swelling during a  calendar year. We  refer to~\cite[Sections~1
and 2]{Charpentier_2022} for  an excellent introduction to  drought events and
their economic  consequences.  In  a nutshell,  the clay  present in  the soil
alternatively shrinks  and swells in  dry and humid conditions.   This creates
instabilities  and generates  cracks in  buildings.  The  cracks induce  costs
covered by all private  property insurance policies~\citep{MTES2016}.  Because
90\%  of  the  French  natural  disasters insurance  market  is  reinsured  by
\href{https://www.ccr.fr/en/}{\underline{Caisse   Centrale  de   Réassurance}}
(henceforth  abbreviated  as   CCR)~\citep{CCR_rapport2021},  a  public-sector
reinsurer providing cedents operating in  France with coverage against natural
catastrophes and uninsurable risks, it is  the French state that is eventually
exposed.

France has been  facing severe drought events over the  past years.  According
to  CCR ~\citep{CCR_1982-2020},  the  average annual  cost  of drought  events
between 2016 and  2020 is 1.1 billion euros, a  threefold increase relative to
the  2002-2015 period  (in the  aforementioned  reference and  in the  present
study, unless stated otherwise, euros are constant euros).  In this light, the
recent cycle  of extremely intense  drought events raises two  questions: will
climate  change perpetuate  this pattern  \citep{Bradford_2000, Iglesias_2019}
and, if so, what cost will the French state incur?

In a  long-term perspective, CCR has  studied the impact of  climate change on
the damages caused  by natural disasters based on  the Intergovernmental Panel
on      Climate       Change      (IPCC)      scenarios       RCP~4.5      and
RCP~8.5~\citep{CCR_climat2015,CCR_climat2018}.      Resorting    to     ARPEGE
simulations of  the climate  in 2050 provided  by Météo-France,  CCR simulated
damages in  France in 2050  and concluded that the  annual cost in  2050 could
increase, depending on  the scenario, by 3\% (under scenario  RCP~4.5) or 23\%
(under scenario RCP~8.5). Unfortunately, the  latter is more likely today than
the former.

In a short- to middle-term perspective, forecasting the cost of drought events
in France  is an  important task for  CCR.  Due to  intricacies of  the French
legal  framework \citep[known  as the  natural disasters  compensation scheme,
see][Section~2.1]{Charpentier_2022}, the  cost of a  drought event on  a given
year  is the  overall  aggregate  cost across  the  cities  that obtained  the
government declaration of  natural disaster for a drought event  that year. We
stress that we used and will use from now on the word city as a translation of
the  French  word \textit{commune},  a  level  of administrative  division  in
France.  We  use the  word city irrespective  of the  \textit{commune}'s size,
which can vary  widely from small hamlets  to large cities. Going  back to the
forecasting task, it will be carried  out several times every year because, as
months  goes by,  more  relevant  information is  accrued.   At  first, it  is
necessary  to predict  which cities  will make  a request  for the  government
declaration of natural disaster for a drought event. Later on it is known that
some cities did make the request but it is still necessary to make predictions
for  the others.   Later  still it  is  known exactly  which  cities made  the
request.  Note that  once a request is  made, there is no  uncertainty for CCR
about  whether or  not  the city  will obtain  the  government declaration  of
natural disaster for  a drought event.  Therefore CCR  currently addresses two
sub-problems  separately: sub-problem~1  consists in  predicting which  cities
will make a  request for the government declaration of  natural disaster for a
drought  event; sub-problem~2  consists in  predicting the  cost of  a drought
event for  those cities  that obtained the  government declaration  of natural
disaster for a  drought event.  In this study, we  focus on sub-problem~2.  On
the   contrary,   \citep{Charpentier_2022,heranval_2022}   address   the   two
sub-problems as one single problem.  We acknowledge that the problem we tackle
is therefore less challenging than theirs.

To the  best of our knowledge,  \citep{Charpentier_2022,heranval_2022} are the
only two  references available  about the  prediction of  the cost  of drought
events.  The  studies conducted by  insurance companies are  confidential.  In
\citep{Charpentier_2022}, the authors use  Generalized Linear Models (GLM) and
tree-based  machine  learning  algorithms   (variants  of  the  random  forest
algorithm). For a given drought event, for each city, the number of claims and
the  average  cost are  predicted,  then  a  city-specific predicted  cost  is
obtained  by multiplying  these  two  numbers.  The  overall  cost is  finally
estimated   by    the   sum    of   all    the   city-specific    costs.    In
\citep{heranval_2022},  the authors  use  penalized GLM  and machine  learning
algorithms (random  forests and  extreme gradient  boosting) to  predict which
cities will make a request for  the government declaration of natural disaster
for a drought event.  For a given  drought event, for each city susceptible to
make a  request, they  then use a  common linear regression  model to  map the
number  of houses  to  a  city-specific cost.   The  overall  cost is  finally
estimated by the sum of these city-specific costs.

In the present study,  we develop and apply a new  methodology to forecast the
cost     of      a     drought      event     in     France.       Like     in
\citep{Charpentier_2022,heranval_2022},  we  exploit  the Soil  Wetness  Index
(SWI) as  a drought indice \citep[it  is referred to as  the Standardised Soil
Water Index by][]{Charpentier_2022}.  Moreover, like~\citet{Charpentier_2022},
we  also  use  sequential  cross-validation  to take  into  account  the  time
dependence    structure    in    our     data    set.     In    contrast    to
\citep{Charpentier_2022,heranval_2022}, we rely on a richer description of the
cities  which  we  obtained  by  data enrichment  (more  details  to  follow).
Finally, we make predictions based on  an aggregation strategy that adapts the
canonical     Super     Learning     methodology     to     our     framework~
\citep{SL2007,Benkeser2018}.   We  call  our   algorithm  the  One-Step  Ahead
Sequential Super Learner (OSASSL).  Its  theoretical analysis reveals that the
algorithm  can  efficiently   learn  from  our  data   set,  a  \textit{short}
time-series whose  time-specific observations  consist of  a large  network of
slightly dependent data~\citep{osasl}.

In Section~\ref{sec:data}, we  present the data that we collected  and used in
this  study.    In  Section~\ref{sec:osasl},   we  give  a   brief  historical
perspective  of  the concept  of  aggregation  and  describe the  OSASSL.   In
Section~\ref{sec:application}, we  expose and comment  on the results  that we
obtain,  notably  assessing how  the  covariates  used  to predict  the  costs
influence  the  predictions.    In  Section~\ref{sec:discussion},  we  discuss
directions for future work.

\section{Data}
\label{sec:data}

We merge several data  sets into a master data set.  The  merged data sets are
either  provided  by  CCR's  cedents  or   are  collected  by  us  from  other
sources. They contribute different kinds of information.

Of note, in the rest of  this study, France refers to \textit{Metropolitan} or
\textit{Mainland} France. Drought  events are not a threat  in Overseas France
(essentially because there is little clay in these parts of the country).

\subsection{Data provided by CCR's cedents}

Ninety percent of  the French natural disasters insurance  market is reinsured
by CCR~\citep{CCR_rapport2021}.   Contractually, its cedents must  share their
portfolios and  claims data.   Over the  years, CCR has  thus gathered  a vast
collection of accurate localizations and  characteristics of insured goods and
claims data.  From 1990 to present, the collection covers roughly 22\% to 97\%
of    the    overall     cost    of    all    the     French    claims    (see
Section~\ref{subsubsec:costs} and Figure~\ref{graph:exhaustivite}).

\subsection{Data garnered from other sources}

The data set, based so far on  data provided by cedents only, is then enriched
with  data from  four trusted  public  organizations that  collect, share  and
analyze information  about the French  economy and people  (National Institute
for Statistical  and Economic Studies, Insee),  geography (Geographic National
Institute,  IGN), geology  (French  Geological Survey,  BRGM) and  meteorology
(Météo-France).  The new features supplementing  the description of the cities
are seismic and climatic zones, clay shrinkage-swelling hazards, tree-coverage
rate, area, population and years of construction.  Lastly, we benefit from the
Soil  Wetness  Index  (SWI)  as   described  in  \citep{Dirmeyer1999}  and  in
\href{https://donneespubliques.meteofrance.fr/client/document/doc_swi_catnat_277.pdf}{\underline{this
    document}} made available by Météo-France.

\subsection{City-level data processing}
\label{subsec:city:level:data}

Some data  are available  at the  house-level (namely, the  cost of  claim and
insured sum),  but most are not.   In particular, the pivotal  SWI variable is
available at a $8 \times 8$  km$^2$ resolution, while the 90\%-quantile of the
French cities area  is 30 km$^2$.  Consequently,  we choose to work  at a city
level and thus  aggregate the features that have a  higher resolution. Details
follow.

\subsubsection{On the city-level costs of drought events}
\label{subsubsec:costs}

The cost of the damages in a city  caused by a drought event (what will be our
response variable)  is unknown.   However, on  the one  hand the  overall cost
across France is  estimated (in both current and constant  euros) by actuarial
studies conducted by CCR and, on the  other hand, we know the costs \textit{of
  those}  claims filled  in the  claims data  provided by  the cedents  which,
unfortunately, only represent a fraction of all the claims.

Provisional city-specific costs are computed  by aggregating by city the costs
filled in the claims data provided  by the cedents.  Because these claims data
are not  exhaustive, the  sum of  all the  provisional city-specific  costs is
smaller than the estimated overall  cost.  The (final) city-specific costs are
proportional to the provisional city-specific costs in such a way that the sum
of all  the (final) city-specific costs  equals the estimated overall  cost in
constant euros.

\begin{figure}
  \centering \includegraphics[scale=0.8]{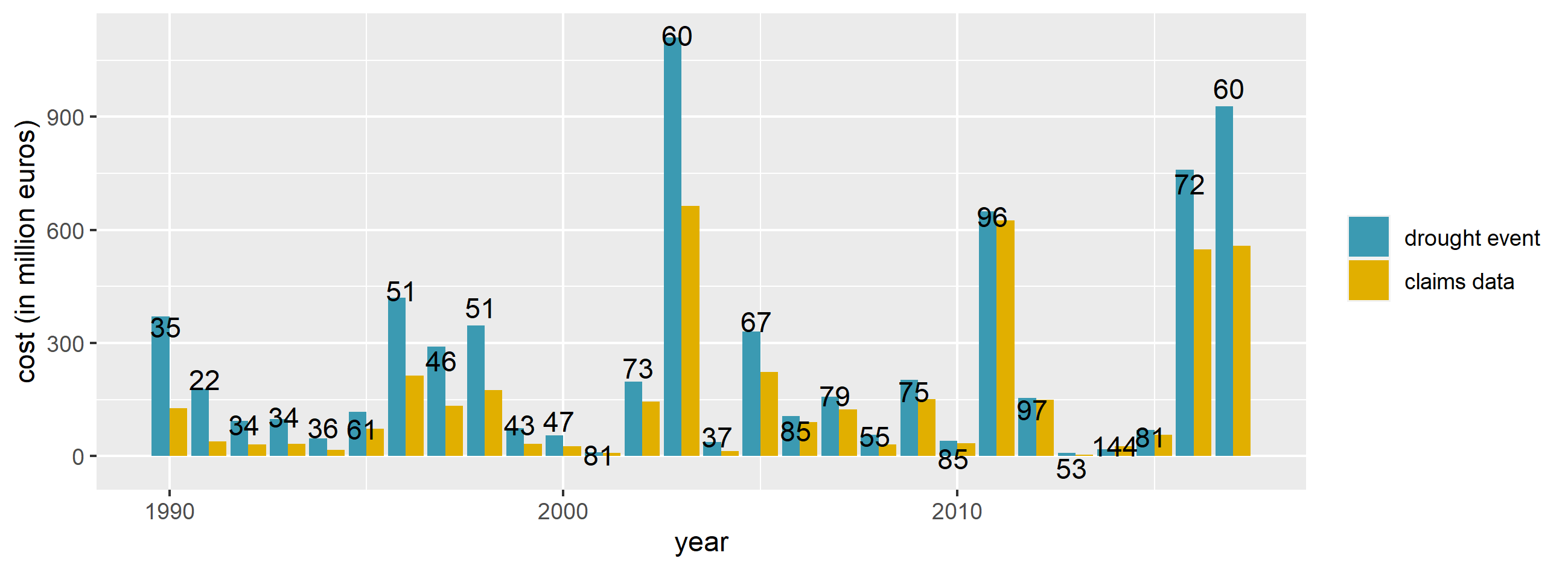}
  \caption{Estimated overall costs of drought  events across France (blue) and
    provisional city-specific costs obtained by aggregating the costs of those
    claims filled  in the claims data  provided by the cedents  (yellow).  The
    ratios  of the  latter to  the  former range  between 22\%  and 97\%  (the
    144\%-ratio corresponds to an exceptional year where the estimated overall
    cost is  even smaller than  the very small  aggregated cost of  the claims
    data). In this figure we use current euros. Source: CCR.}
  \label{graph:exhaustivite}
\end{figure}

Figure~\ref{graph:exhaustivite}  illustrates the  gaps  between the  estimated
overall  costs across  France and  the  sum of  the provisional  city-specific
costs. The ratios of the latter to the former range from 22\% to and 97\% (the
144\%-ratio corresponds  to an  exceptional year  where the  estimated overall
cost is even smaller than the very small aggregated cost of the claims data).

\subsubsection{On the city-level SWI}

For every year and  every city, we derive a collection  of 36 city-level SWIs,
one for each ten-day period (a \textit{décade} in French) that make up a year.
Each of these 36  SWIs is the convex average of the  corresponding SWIs of the
$8 \times  8$ km$^2$ squares  that overlap the  city's area.  The  weights are
proportional to the areas of the intersections.

We  use  the  city-level  SWIs  to build  a  rich  collection  of  SWI-related
covariates.

Because the effects of a drought event can build up slowly, for every year $t$
and every  city, we concatenate the  $3 \times 36$ ten-day  city-level SWIs of
years $t$,  $(t-1)$ and $(t-2)$.  We  also add the minima,  means and standard
deviations  of the  36 ten-day  city-level SWIs  computed separately  over the
years $t$, $(t-1)$ and $(t-2)$.

In addition, for every year $t$ and every city, we compute and concatenate the
mean SWI of all ten-day periods  from April to September for \textit{(a)} year
$t$ alone, \textit{(b)}  years $t$ and $(t-1)$, \textit{(c)}  years $t, (t-1)$
and $(t-2)$. The  period April to September corresponds to  the dry season, as
opposed to the period October to March, which corresponds to the wet season.

Moreover,  for  each  quarter~$1\leq   q\leq  4$  (January-March,  April-June,
July-September, October-December), for every year $\tau$ between 1959 and 2017
and every  city $\alpha$, we  compute the  average city-level SWI,  denoted by
$\barSWI_{q,\tau,\alpha}$, and form the four cumulative distribution functions
$\hat{F}_{q}$      associated       to      the      four       data      sets
$\{\barSWI_{q,\tau,\alpha} : 1959  \leq \tau \leq 2009,  \alpha\}$.  Then, for
every year  $1990 \leq t\leq  2017$ and every city  $\alpha$, we also  add the
$3\times                            4$                           probabilities
$\hat{F}_{q}(\barSWI_{q,t,\alpha}),       \hat{F}_{q}(\barSWI_{q,t-1,\alpha}),
\hat{F}_{q}(\barSWI_{q,t-2,\alpha})$ ($q=1,\ldots, 4$).

\subsubsection{On the city-level description}

For  every year,  each city  is described  by a  collection of  covariates.  A
city's multi-faceted  description attempts  to capture  all the  city's traits
that, beyond  the city-level  SWIs presented in  the previous  subsection, can
explain the cost of a possible drought event. It contains:
\begin{itemize}
\item The year $t$.
\item  The city's  area, average  altitude, climatic  zone, seismic  zone, and
  proportions of  surface with a  ``tree-coverage'' greater than  10\%.  Here,
  climatic        zone        is        a        five-category        variable
  \href{https://www.legifrance.gouv.fr/loda/article_lc/LEGIARTI000026910138}{\underline{attributed
      by the French State}} to each French department (one of the three levels
  of government under  the national level, between  the administrative regions
  and the communes; metropolitan France  counts 96 departments).  Seismic zone
  is  a  four-category  variable  attributed   to  each  city  by  the  French
  \href{https://www.legifrance.gouv.fr/codes/id/LEGIARTI000022959104/2011-05-01}{\underline{Code
      de l'environnement}}.   As for ``tree-coverage'', obtained  from IGN, it
  corresponds   to  any   of   the   17  types   of   terrain  documented   in
  \cite[Section~11.4, page~183, variable NATURE]{IGN2021}.
\item The number  of inhabitants, (estimated) number of  houses located within
  the city's  limits, house  density, defined  as the ratio  of the  number of
  houses to the city's area, and proportions of buildings built prior to 1949,
  between 1950 and  1974, between 1975 and 1989, and  after 1989.  The numbers
  of  inhabitants come  from~\citep{COGugaison}  (an  \texttt{R} package  that
  integrates                  data                   from                  the
  \textit{\href{https://www.insee.fr/fr/information/2560452}{\underline{Code
        Officiel  Géographique  de  l'Insee}}).   The  number  of  houses  are
    estimated  by  CCR  based  on census  data~\citep{INSEE2000}  and  on  the
    \href{https://www.data.gouv.fr/fr/datasets/base-sirene-des-entreprises-et-de-leurs-etablissements-siren-siret/}{\underline{SIRENE
        database}} obtained  from Insee.   They are confidential.   The citys'
    areas     are     found    in~\cite[Section~6.1,     page~12,     variable
    SUPERFICIE]{GEOFLA}.   As  for  the  proportions of  buildings,  they  are
    computed   based    on   data    compiled   by   Insee    and   documented
    in~\citep{INSEE2000}.  Note that  the thresholds 1949, 1974  and 1989 used
    to describe the age of the housing stock were fixed by Insee.}
\item The proportions of the houses located within the city's limits that fall
  in  each  of  the  four clay  shrinkage-swelling  hazards  categories.   The
  four-category  clay  shrinkage-swelling  hazard  variable  is  defined,  and
  obtained from, BRGM~\citep{MI2019}. Its resolution is fine enough for a city
  to fall in more than one category.
\item The (estimated)  insured sum corresponding to the  houses located within
  the city's limits, and the average house  value, defined as the ratio of the
  aforementioned insured  sum to the  number of  houses. The insured  sums are
  evaluated by  CCR based on data  from Insee and portfolios  data provided by
  CCR's   cedents.   The   insured  sums   are  confidential.    We  use   the
  \href{https://www.outils.ffbatiment.fr/federation-francaise-du-batiment/le-batiment-et-vous/en_chiffres/indices-index/Chiffres_Index_FFB_Construction.html}{\underline{Indice
      de la Fédération Française du Bâtiment}} to account for inflation.
\item  Six indicators  of whether  or  not the  city  made a  request for  the
  government declaration of natural disaster for  a drought event on years $t$
  to  $(t-5)$, and  six indicators  of whether  or not  the city  obtained the
  government declaration of natural disaster for  a drought event on years $t$
  to $(t-5)$.  We  could derive these indicators because,  being the secretary
  of the Commission Interministérielle  Catastrophe Naturelle, CCR has access,
  every  year, to  the  list of  all  cities  that made  a  request for  (and,
  possibly, obtained)  the government  declaration of  natural disaster  for a
  drought event.  The data are publicly available~\citep{CCR_arretes2022}.
\end{itemize}
In addition, the city's description contains:
\begin{itemize}
\item  the cumulated  city-level costs  computed across  the years  $(t-1)$ to
  $(t-5)$;
\item the  mean and  median city-level  costs of  the drought  events computed
  across  the  years  $(t-1)$  to  $(t-5)$ and  all  cities  within  the  same
  department.
\end{itemize}

The   city's   description   is   finally   enriched   with   \textit{compound
  covariates}. The compound covariates have a similar form. For every year $t$
and each city $\alpha$, for a given covariate $C_{h,\alpha,t}$ defined for all
houses $h$  within the city's limits  (and in the portfolios  data provided by
the       cedents),       we       compute       the       weighted       mean
$\sum_{h} s_{h,\alpha,t} \times  C_{h,\alpha,t}/\sum_{h} s_{h,\alpha,t}$ where
$s_{h,\alpha,t}$ is  the (estimated) insured  sum of house $h$  located within
$\alpha$'s limits on year $t$. Here $C_{h,\alpha,t}$ can be:
\begin{enumerate}
\item\label{enum:1} the  mean of all the  $t$-specific 36 ten-day SWIs  of the
  $8 \times 8$km$^2$ square which contains house $h$;
\item\label{enum:2} the level of  the clay shrinkage-swelling hazard localized
  at $h$ (does not depend on $t$);
\item\label{enum:3}  the ground  slope localized  at $h$  (does not  depend on
  $t$);
\item     the     three     products     \eqref{enum:1}$\times$\eqref{enum:2},
  \eqref{enum:1}$\times$\eqref{enum:3},
  \eqref{enum:1}$\times$\eqref{enum:2}$\times$\eqref{enum:3}.
\end{enumerate}
Moreover,   for  every   year  $t$   and   every  city   $\alpha$,  for   each
$C_{h,\alpha,t}$ among  \eqref{enum:1}, \eqref{enum:2} and  \eqref{enum:3}, we
also     add      the     30     29-quantiles     of      the     data     set
$\{s_{h,\alpha,t} \times C_{h,\alpha,t} : h\}$  (where $h$ ranges over the set
of  houses $h$  within $\alpha$'s  limits).  Overall,  the city's  description
consists of a slightly fewer than 400 covariates.

\section{The One-Step Ahead Sequential Super Learner}
\label{sec:osasl}

The  One-Step Ahead  Sequential Super  Learner (OSASSL)  adapts the  canonical
Super  Learning  methodology,  one  among many  strategies  to  aggregate  the
predictions of several predictors.   In Section~\ref{subsec:stacking}, we give
a brief historical perspective of the  concept of aggregation and describe the
canonical Super  Learning methodology in  the simple context where  one learns
from independent and identically distributed data.  In Section~\ref{subsec:SL}
we present OSASSL and succinctly review its theoretical performance.  Finally,
in Section~\ref{subsec:pen}, we explain how OSASSL can be used to forecast the
cost of drought events.

\subsection{Aggregation strategies}
\label{subsec:stacking}

The idea  of aggregating  several estimation strategies  to take  advantage of
their respective  strengths emerged in  the 1990s. The principle  of ``stacked
generalization''   was  introduced   by  \cite{wolpert1992stacked}.    Stacked
generalization   consisted  in   combining   several  lower-level   predictive
algorithms  into a  higher-level  meta-algorithm with  the  aim of  increasing
predictive accuracy.  Later, \cite{breiman1996stacked} showed how ``stacking''
can be used to improve predictive accuracy in a regression context, and how to
impose constraints  on the higher-level  algorithm in order to  achieve better
predictive performance.  Since then, stacking has been evolving into a variety
of    methods    among    which    is    the    canonical    Super    Learning
methodology~\citep{SL2007,SLchapter}.

Related literature introduces and discusses the concepts of boosting, bagging,
random
forests~\citep{freund1995boosting,breiman1996bagging,amit1997shape,breiman2001random},
and  robust  online  aggregation  (also  known  as  prediction  of  individual
sequences or prediction with expert advice)~\citep{Warmuth1994,Lugosi2006}.  A
Bayesian  perspective  on  ``model averaging''  was  concomitantly  introduced
by~\cite{raftery1999}.   All these  aggregation strategies  have thrived  both
theoretically and in applications. 

We revealed in the  introduction that we chose to focus  on Super Learning. We
have been asked to justify this choice  and to explain why the methodology has
advantages  over  others.   Let  us  stress  first  that  none  of  the  above
methodologies would have applied  off-the-shelf with theoretical guarantees of
performance because of the complex dependence structure induced in the data by
the spatial and temporal nature of drought events.  Being well versed in Super
Learning,  from both  the  theoretical and  applied  viewpoints, we  naturally
favored  this  methodology.   We  believe  that  Sections~\ref{subsec:SL}  and
\ref{sec:application} show that this was a good decision from both theoretical
and  applied  viewpoints.   Furthermore,  in the  application,  we  use  Super
Learning as an aggregation strategy to compare and combine several lower-level
algorithms which are themselves Super Learners.  We thus use Super Learning as
an   aggregation  strategy   to  compare   and  combine   several  aggregation
strategies. So,  although we focus  on Super Learning,  our scope is  far from
being narrow.

In brief, the canonical Super Learning methodology is a general methodology to
learn a feature of  the law of the data identified  through an \textit{ad hoc}
risk  function  by  relying  on  a library  of  (low-level)  algorithms.   The
algorithms either compete (discrete Super Learning methodology) or collaborate
through   a   (higher-level)   meta-algorithm   (continuous   Super   Learning
methodology), with  a cross-validation scheme determining  the best performing
algorithm or  combination of  algorithms, respectively.   We refer  the reader
to~\citep{NB2018}  for  a   gentle  introduction  to  Super   Learning  and  a
step-by-step development  of two examples  to illustrate concepts  and address
common concerns.

In  the  simpler  case  where  one learns  from  independent  and  identically
distributed  data, one  often implements  a $V$-fold  cross-validation scheme:
first, the  data set  is split  into $V$  groups of  roughly equal  sizes (the
``folds''); second, every algorithm is trained  and tested $V$ times, once for
each fold, with each fold being used  for testing after the algorithm has been
trained using all the other folds; third, the cross-validated (empirical) risk
of  the  algorithm  is  defined  as  the  average  of  the  $V$  fold-specific
(empirical) risks obtained by testing. In the present study, however, we learn
from a (short) time-series (with time-specific observations consisting of many
dependent data-structures) and thus cannot rely on a $V$-fold cross-validation
scheme.    Instead,  like~\cite{Benkeser2018},   we  rely   on  a   sequential
cross-validation scheme:  sequentially at each  time $t$, for  each algorithm:
all data till time $(t-1)$ are used for training and the $t$-specific data are
used for testing; the $t$-specific cross-validated (empirical) cumulative risk
of the algorithm is defined as  the average of the $\tau$-specific (empirical)
risks (where $\tau$ ranges between 1 and $t$) obtained by testing. Remarkably,
the sequential  cross-validation scheme  can neglect the  dependence structure
within each time-specific  observation (in particular, it is  not necessary to
cross-validate spatially).  Details follow.

\subsection{Presentation and theoretical performance of OSASSL}
\label{subsec:SL}

In~\citep{osasl}, building  upon the canonical Super  Learning methodology, we
developed and  studied OSASSL, an  algorithm designed to learn  a (stationary)
feature  of  the  law  of  a  \textit{short}  time-series  with  time-specific
observations  consisting  of  \textit{many}  dependent  data-structures.   The
analysis   of  OSASSL   reveals  that   it  manages   to  make   up  for   the
\textit{shortness} of the time-series thanks  to the \textit{manyness} of each
time-specific  observation   provided  that  the  latter   are  only  slightly
dependent.  Moreover,  as already  stressed, OSASSL does  not require  that we
model and take into account the dependence structure within each time-specific
observation. Here, we present an instance  of the OSASSL specifically built to
forecast the cost of drought events.

We let  $(\bar{O}_{t})_{t\geq 1}$ denote  the time-series that  formalizes the
time-series    described   in    section~\ref{sec:data}.     At   each    time
$t  \in  \mathbb{N}^{*}$,  $\bar{O}_{t}$   consists  of  a  finite  collection
$(O_{\alpha,  t})_{\alpha \in  \calA}$ of  $(\alpha,t)$-specific observations,
where  each  $\alpha   \in  \calA$  represents  a  French   city.   For  every
$\alpha      \in      \calA$,      $O_{\alpha,     t}$      decomposes      as
$O_{\alpha,t}  := (Z_{\alpha,t},  X_{\alpha,t}, Y_{\alpha,t})\in  \calZ \times
\calX \times [0,B] =: \calO$ where $X_{\alpha,t}\in\calX$ is the collection of
covariates describing the city $\alpha$ on year $t$ (including an indicator of
whether  or  not the  city  obtained  the  government declaration  of  natural
disaster for  a drought event),  $Z_{\alpha,t}\in\calZ$ is the  city-level SWI
describing the  drought event  that year, and  $Y_{\alpha,t} \in[0,B]$  is the
city-specific cost  of the drought event  that year (known to  take its values
between 0 and a  constant $B$). By convention, $Y_{\alpha,t} =  0$ if the city
did not  obtain the government declaration  of natural disaster for  a drought
event.

We       assume        that       the       mean        conditional       cost
$\theta^{\star}  :  (x,z)  \mapsto  \Exp\left[Y_{\alpha,t}|X_{\alpha,t}  =  x,
  Z_{\alpha,t} = z\right]$ does not depend on $(\alpha,t)$ or, in other terms,
that it is a stationary feature  of the law of $(\bar{O}_{t})_{t\geq 1}$. This
is   the   case  if,   given   a   time-specific  city-description   and   SWI
$(X_{\alpha,t}, Z_{\alpha,t})$,  the mechanism  that produces  a cost  after a
drought event conditionally on  $(X_{\alpha,t}, Z_{\alpha,t})$ does not depend
on $(\alpha,t)$, that  is, remains constant throughout time  and France. Under
this stationarity assumption, we can use the estimator of the mean conditional
cost to make predictions  at any $(x, z)$ provided that $(x,  z)$ falls in the
domain  of  the  observed  $(X_{\alpha,t} ,  Z_{\alpha,t})$.   Naturally,  the
scarcer  the available  information around  $(x,  z)$, the  less reliable  the
prediction.  Moreover, if $(x, z)$ falls  outside the domain, then, although a
prediction may  be made  nonetheless, it  cannot be trusted.   So, in  view of
climate change,  not-too-distant-future projections  of drought events  can be
made.

In this study the OSASSL is  a meta-algorithm that learns the mean conditional
cost  $\theta^{\star}$   from  $(\bar{O}_{t})_{t\geq   1}$  by   stacking  the
estimators of $\theta^{\star}$  provided by a user-supplied  collection of $J$
algorithms $\widehat{\theta}_{1}, \ldots, \widehat{\theta}_{J}$.  At each time
$t   \geq    1$,   every    algorithm   $\widehat{\theta}_{j}$    trained   on
$\bar{O}_{1},  \ldots, \bar{O}_{t}$  outputs  an  estimator $\theta_{j,t}$  of
$\theta^{\star}$.   Every algorithm  $\widehat{\theta}_{j}$ is  constrained so
that $\theta_{j,t}(x,z)=0$  if $x$ conveys  the information that the  city did
not obtain the government declaration of natural disaster for a drought event.
The OSASSL  selects the best  algorithm indexed by $\widehat{j_t}$  defined as
the minimizer of the empirical average cumulative risks,
\begin{equation}
  \label{eq:oSL}
  \widehat{j}_{t} \in \argmin_{1 \leq j \leq J} \widehat{R}_{j,t},
\end{equation}
where
\begin{equation}
  \label{eq:erisk}
  \widehat{R}_{j,t}
  :=   \frac{1}{t|\calA|}   \sum_{\tau=1}^{t}   \sum_{\alpha   \in   \calA}
  \left[Y_{\alpha,\tau}        -         \theta_{j,        \tau-1}(X_{\alpha,\tau},
    Z_{\alpha,\tau})\right]^{2}.  
\end{equation}
Interestingly, the OSASSL is an online algorithm if each of the $J$ algorithms
$\widehat{\theta}_{1}, \ldots, \widehat{\theta}_{J}$ is  online, that is, such
that the  making of $\theta_{j,t}$  consists in an update  of $\theta_{j,t-1}$
based on newly accrued data $\bar{O}_{t}$.

The $t$-specific measure of performance  of each $\widehat{\theta_{j}}$ is the
unknown quantity
\begin{equation}
  \label{eq:risk}
  \widetilde{R}_{j,t}
  := \frac{1}{t|\calA|} \sum_{\tau=1}^{t} \sum_{\alpha       \in       \calA}
  \Exp\left\{\left[Y_{\alpha,\tau} - \theta_{j, \tau-1}(X_{\alpha,\tau}, 
  Z_{\alpha,\tau})\right]^{2} \middle| \bar{Z}_{\tau}, F_{\tau-1}\right\}
\end{equation}
where $F_{t}$ is  the history generated by  $\bar{O}_{1}, \ldots, \bar{O}_{t}$
(by  convention, $F_{0}  =  \emptyset$).   It takes  the  form  of an  average
cumulative risk  conditioned on the  sequence $(\bar{Z}_{t})_{t \geq  1}$ with
$\bar{Z}_{t} = (Z_{\alpha,t})_{\alpha \in  \calA}$.  The $t$-specific oracular
meta-algorithm is indexed  by the oracular $\widetilde{j}_{t}$  defined as the
minimizer
\begin{equation}
  \label{eq:oracle}
  \widetilde{j}_{t} \in \argmin_{1 \leq j \leq J} \widetilde{R}_{j,t},
\end{equation}
which,  like  each  $\widetilde{R}_{j,t}$,  is   unknown  to  us.   Note  that
$\widehat{R}_{j,t}$  estimates $\widetilde{R}_{j,t}$  and that  \eqref{eq:oSL}
mimics \eqref{eq:oracle}.

The  theoretical analysis  hinges  on a  key-assumption  about the  dependence
structure  in   the  time-series  $(\bar{O}_{t})_{t  \geq   1}$.   We  exploit
conditional dependency graphs to model the amount of conditional independence.
Specifically, we assume the existence of  a graph $\calG$ with vertex and edge
sets $\calA$ and $\calE$  such that if $\alpha \in \calA$  is not connected by
any   edge  $e\in\calE$   to  any   $\alpha'\in\calA'  \subset   \calA$,  then
$O_{\alpha,t}$         is         conditionally         independent         of
$(O_{\alpha',t})_{\alpha' \in \calA'}$ given $F_{t-1}$ and $\bar{Z}_{t}$. Then
what matters  is the connectedness of  the graph, as reflected  by its degree,
$\deg(\calG)$,  which equals  1  plus the  largest number  of  edges that  are
incident to a vertex in $\calG$. Finally, let us emphasize that the dependency
graph $\calG$ plays no role in the OSASSL's characterization and training.  In
other words,  as stated at  the beginning of  this section, we  can altogether
neglect the intricate spatial  dependence within each $\bar{O}_{t}$.  However,
the key-assumption is pivotal in the algorithm's theoretical analysis.

The performance of $\widehat{j}_{t}$ as an estimator of $\widetilde{j}_{t}$ is
expressed in  terms of a comparison  of the excess average  cumulative risk of
the  former to  the  excess  average cumulative  risk  of  the latter.   Under
additional  mild  assumptions   \citep[][corollary~2]{osasl}  there  exists  a
decreasing function  $C :  \bbR_{+}^{*} \to \bbR_{+}^{*}$  such that,  for any
$\varepsilon > 0$,
\begin{equation}
  \label{eq:cor}
  \Exp           \left[            \underbrace{\widetilde{R}_{\widehat{j}_t,t}           -
      \widetilde{R}_{t}(\theta^{\star})}_{\text{excess risk of } \widehat{j}_t}
    - (1 + \varepsilon) 
    \Big(\underbrace{\widetilde{R}_{\widetilde{j}_t,                 t}                 -
      \widetilde{R}_{t}(\theta^{\star})}_{\text{excess     risk     of     }
      \widetilde{j}_t}\Big)   \right]   \leq   C(\varepsilon)   \frac{\log
  (J\log(\calI^{2}))}{\calI^{2}} 
\end{equation}
where $\calI^{2}$  grows like the amount  of information available and  can be
equal   to   either   $t$   or   $|\calA|/(t\deg(\calG))$.    If   the   ratio
$|\calA|/\deg(\calG)$  is  sufficiently  large  (both in  absolute  terms  and
relative to $t$), then the  oracular inequality \eqref{eq:cor} is sharper when
$\calI^{2} = |\calA|/(t\deg(\calG))$  than when $\calI^{2} =  t$. This reveals
that the OSASSL  can leverage a large ratio $|\calA|/\deg(\calG)$  in the face
of a small $t$.

In  the  application,  $t  \approx  25$, $|\calA|  \approx  36,000$.   As  for
$\deg(\calG)$, it is much harder to assess a meaningful value. In this regard,
it is relevant to recall that,  in 2019, France had around $1,000$ federations
of cities, each regrouping 30 cities  on average. Furthermore, we computed the
number of neighboring  cities for each city.  The quantiles  and mean of these
numbers are  reported in Table~\ref{tab:nb:neigh:cities}.  In  particular, the
city with the largest number of neighboring cities (Paris) has 29 of them.

\begin{table}
  \centering
  \begin{tabular}{ccccccc}\hline
    min. & 1st qu. & median & mean & 3rd qu. & 99\%-qu. & max\\\hline
    0 & 5 & 6 & 5.96 & 7 & 11 & 29\\\hline
  \end{tabular}
  \caption{Quartiles,  99\%-quantile and  mean of  the numbers  of neighboring
    cities  in France  in 2019.   Although the  maximum cannot  be interpreted
    literally  as $\deg(\calG)-1$,  it nevertheless  gives a  sense of  what a
    meaningful value of $\deg(\calG)$ could be.}
  \label{tab:nb:neigh:cities}
\end{table}

\subsection{Forecasting the cost of drought events}
\label{subsec:pen}

The OSASSL presented in Section~\ref{subsec:SL}  is designed to learn the mean
conditional  cost $\theta^{\star}$  from $(\bar{O}_{t})_{t  \geq 1}$.  At each
time    $t    \geq    1$,    it    outputs    the    $t$-specific    estimator
$\theta_{\widehat{j}_{t},  t}$.   This estimator  can  be  evaluated at  every
$(X_{\alpha, t+1}, Z_{\alpha,t+1})$ ($\alpha \in \calA$) and we use the sum
\begin{equation*}
  \sum_{\alpha\in\calA}    \theta_{\widehat{j}_{t},     t}(X_{\alpha,    t+1},
  Z_{\alpha,t+1}) 
\end{equation*}
to  predict  the  cost  of  the  drought  event  at  time  $(t+1)$,  that  is,
$\sum_{\alpha \in \calA} Y_{\alpha,t+1}$.

\section{Application}
\label{sec:application}

This section discusses the practical implementation, training and exploitation
of   the   OSASSL   presented    and   studied   in   Section~\ref{sec:osasl}.
Section~\ref{sec:appli:library}  describes the  collection  of $J$  algorithms
$\widehat{\theta}_{1},              \ldots,             \widehat{\theta}_{J}$.
Section~\ref{sec:appli:training}   explains  how   the   OSASSL  is   trained.
Section~\ref{sec:appli:results} presents the results and comments upon them.

\subsection{Implementing two OSASSLs}
\label{sec:appli:library}

We deploy  two meta-algorithms taking  the form  of OSASSLs, the  discrete and
continuous overarching Super  Learners.  Both rely on the same  library of $J$
algorithms  $\widehat{\theta}_{1}, \ldots,  \widehat{\theta}_{J}$.  These  $J$
algorithms are  themselves OSASSLs either  in the strict  or in a  loose sense
(more details to follow).

\subsubsection{Penalization}

Because our ultimate goal is to forecast the cost of the latest drought event,
we made  the decision to  rely on  a penalized version  of $\widehat{R}_{j,t}$
\eqref{eq:erisk}, by substituting
\begin{equation}
  \label{eq:erisk:pen}
  \widehat{R}_{j,t}     +      \frac{0.05}{t}     \sum_{\tau      =     1}^{t}
  \Big(\underbrace{\sum_{\alpha\in \calA} Y_{\alpha,\tau}}_{\text{actual cost}} -
  \underbrace{\sum_{\alpha\in\calA}
    \theta_{\widehat{j}_{\tau-1},   \tau-1}   (X_{\alpha,   \tau},
    Z_{\alpha, \tau})}_{\text{predicted cost}}\Big)^{2} 
\end{equation}
for  $\widehat{R}_{j,t}$  (we recall  that  $\theta_{j,t}$  is the  output  of
$\widehat{\theta}_{j}$ trained on $\bar{O}_{1},  \ldots, \bar{O}_{t}$ and that
$\widehat{j}_{t}$   is   defined   in~\eqref{eq:oSL}).   Observe   that   each
$t$-specific   penalization  term   equals   0.05  times   the  average   over
$1\leq  \tau \leq  t$ of  the $\tau$-specific  squared difference  between the
actual cost  of the drought event  (left-hand side summand) and  the predicted
cost      made     by      the     (penalized)      OSASSL     trained      on
$\bar{O}_{1}, \ldots, \bar{O}_{\tau-1}$ (right-hand side summand).  The factor
0.05 was chosen somewhat arbitrarily.

By adding this penalization term, the OSASSL favors the algorithms that better
predict   not   only   the   \textit{city-specific}   costs   but   also   the
\textit{overall}  cost   of  the  next   drought  event.   In   addition,  the
penalization term slightly  dilutes the importance of  the city-specific costs
and,  on the  contrary, reinforces  the importance  of the  overall cost,  the
latter  being   more  dependable   than  the  former   as  we   explained  in
Section~\ref{subsec:city:level:data}.

\subsubsection{The discrete and continuous overarching Super Learners}

Called the  \textit{discrete} overarching Super  Learner, the first  OSASSL is
the algorithm  that, at  time $t\geq 1$,  outputs $\theta_{\widehat{j}_{t},t}$
(using  \eqref{eq:erisk:pen}  instead  of  \eqref{eq:erisk}  as  an  empirical
measure of  the risk).   In words,  at time  $t \geq  1$, the  algorithm whose
penalized empirical average cumulative risk  is the smallest is determined and
the discrete  overarching Super Learner  returns the output of  that algorithm
trained on all data till time $t$.

We also consider a second OSASSL which is defined as a regular OSASSL based on
a  library derived  from $\widehat{\theta}_{1},  \ldots, \widehat{\theta}_{J}$
and   comprising  $J'   =  \mathcal{O}(\varepsilon^{1-J})$   algorithms  where
$\varepsilon>0$        is        a         small        positive        number
($J' = \mathcal{O}(\varepsilon^{1-J})$  means that $J'$ is  upper-bounded by a
constant times $\varepsilon^{1-J}$).  Specifically,  these $J'$ algorithms are
denoted by $\widehat{\boldsymbol{\theta}}_{\pi}$ where  the index $\pi$ ranges
in         an        $\varepsilon$-net         over        the         simplex
$\{x  \in (\bbR_{+})^{J}  : \sum_{j=1}^{J}x_{j}  = 1\}$  (an $\varepsilon$-net
whose cardinality is  $J'$, that is, a  finite subset of $J'$  elements of the
simplex  which  ``approximates''   the  simplex).   For  each   $\pi$  in  the
$\varepsilon$-net,     $\widehat{\boldsymbol{\theta}}_{\pi}$    trained     on
$\bar{O}_{1},   \ldots,  \bar{O}_{t}$   outputs   the  $\pi$-specific   convex
combination   $\sum_{j=1}^{J}    \pi_{j}   \theta_{j,t}$.    The    bound   in
\eqref{eq:cor}           is           still          meaningful           when
$\varepsilon =  \mathcal{O}(\calI^{-1})$.  We refer  to this second  OSASSL as
the \textit{continuous} overarching Super Learner.

\subsubsection{The  discrete and  continuous overarching  Super Learners'
    library of algorithms}

We    now    turn    to    the   description    of    the    $J$    algorithms
$\widehat{\theta}_{1}, \ldots,  \widehat{\theta}_{J}$. All  of them rely  on a
collection                  of                  base                  learners
$\widehat{\calL}_{1}, \ldots, \widehat{\calL}_{K}$.  Some of the base learners
rely  on  linear models  and  their  extensions  (lasso, ridge,  elastic  net,
multivariate adaptive regression splines,  support vector regression).  Others
are tree-based algorithms (CART, random forest, gradient boosting), or rely on
neural  networks.   Others  fall  in  the  category  of  $k$-nearest-neighbors
algorithms  tailored   to  our  study   so  that  the   dissimilarity  between
observations  is  a convex  combination  of  the Kolmogorov-Smirnov  distances
between the  empirical average cumulative distribution  functions mentioned in
Section~\ref{subsec:city:level:data}.    Finally,  some   are  regular   Super
Learners themselves, based on a  selection of the aforementioned base learners
and oblivious  to the temporal ordering  (that is, they rely  on vanilla inner
$V$-fold cross-validation).

Moreover, some of  these base learners are combined  (upstream) with screening
algorithms.   A screening  algorithm is  merely  an algorithm  that selects  a
subset  of the  covariates  deemed relevant  to feed  the  base learners.   In
general, the  selection can  be either deterministic  or data-driven.   In our
study,  we  only  use  deterministic  screening  algorithms  based  on  expert
knowledge.

Overall,  we implement  a collection  of $K=27$  base learners  (including the
variants  obtained  by combining  with  different  screening algorithms).  The
collection       is       shared       by       the       $J$       algorithms
$\widehat{\theta}_{1},  \ldots,  \widehat{\theta}_{J}$  which  differ  in  the
methods they rely on to exploit the base learners.

One of the  method yields a OSASSL precisely as  defined in \eqref{eq:oSL} and
\eqref{eq:erisk}/\eqref{eq:erisk:pen}  where we  substitute  $K$  for $J$  and
$\ell_{j,\tau - 1}$ for $\theta_{j,  \tau-1}$, with $\ell_{j,t}$ the output of
$\widehat{\calL}_{j}$  trained  on  $\bar{O}_{1}, \ldots,  \bar{O}_{t}$.   The
resulting  OSASSL is  an  instance  of discrete  Super  Learner as  previously
described when introducing the first overarching Super Learner.  As we already
explained,         the         library        of         base         learners
$\widehat{\calL}_{1}, \ldots,  \widehat{\calL}_{K}$ can  be extended  using an
$\varepsilon$-net                over               the                simplex
$\{x \in (\bbR_{+})^{K}  : \sum_{k=1}^{K} x_{k} = 1\}$. For  each $\pi$ in the
$\varepsilon$-net,     $\widehat{\boldsymbol{\calL}}_{\pi}$     trained     on
$\bar{O}_{1},   \ldots,  \bar{O}_{t}$   outputs   the  $\pi$-specific   convex
combination   $\sum_{k=1}^{K}  \pi_{k}   \ell_{k,t}$.    Using  the   extended
collection  of base  learners,  the same  method then  yields  an instance  of
continuous Super Learner  as previously described when  introducing the second
overarching Super Learner.

In a similar fashion, we consider several methods to exploit the base learners
$\widehat{\calL}_{1},   \ldots,   \widehat{\calL}_{K}$.   Heuristically,   the
principle is  to learn to  produce a single  prediction based on  the multiple
predictions made by  the base learners once they have  been trained, just like
we  described in  the  previous  paragraph.  Two  natural  and simple  methods
consist in using the average or  the median of the base learners' predictions.
Some methods  rely on  the same method  as in the  previous paragraph  with an
extra penalization term in the definition of the risk (similar to the one used
to define \eqref{eq:erisk:pen} based  on \eqref{eq:erisk}).  The other methods
rely on the lasso, ridge and elastic net algorithms, or on the random forests,
gradient boosting and support vector  regression algorithms.  Finally, some of
the methods  can exploit the  covariates.  Overall, we implement  a collection
$J=50$ algorithms $\widehat{\theta}_{1}, \ldots, \widehat{\theta}_{J}$.

\subsection{Training the discrete and continuous overarching Super Learners} 
\label{sec:appli:training}

At each time $t \geq 1$ we define  a summary of the past based on observations
made during the five previous years.  To  do so, we reserve the data from year
1990  to  year  1994.  This  is  very  relevant  for  two reasons.   First,  a
drought-related  claim can  be the  by-product of  repeated shrinkage-swelling
episodes  over the  years. Second,  a city-level  cost of  a drought  event is
expected to be high  when the city did not benefit  recently from a government
declaration of natural  disaster for a drought event (because  of the possible
accumulation of damages over the years); on the contrary, it is expected to be
low otherwise (because damages may already have been compensated).

For each $t \in \{1995,  \ldots, 1999\}$, we derive $\ell_{1,t-1994}$,~\ldots,
$\ell_{K,t-1994}$.   For  each  $t  \in \{2000,  \ldots,  2005\}$,  we  derive
$\theta_{1,t-1994}$,~\ldots,             $\theta_{J,t-1994}$             using
$\ell_{1,(t-1)-1994}$,~\ldots,  $\ell_{K,(t-1)-1994}$,  and   we  also  derive
$\ell_{1,t-1994}$,~\ldots,         $\ell_{K,t-1994}$.         For         each
$t  \in \{2006,  \ldots, 2017\}$,  we  derive the  discrete overarching  Super
Learner    $\widehat{j}_{t-1994}$    using    $\theta_{1,(t-1)-1994}$,~\ldots,
$\theta_{J,(t-1)-1994}$        (which         rely        themselves        on
$\ell_{1,(t-2)-1994}$,~\ldots,  $\ell_{K,(t-2)-1994}$),  and  we  also  derive
$\theta_{1,t-1994}$,~\ldots,              $\theta_{J,t-1994}$              and
$\ell_{1,t-1994}$,~\ldots,         $\ell_{K,t-1994}$.         For         each
$t \in  \{2006, \ldots, 2017\}$,  the continuous overarching Super  Learner is
derived too.

To this  day, the real  costs and city-level costs  for the years  2018, 2019,
2020 and 2021 are still uncertain.  We thus cannot train our algorithms beyond
the year 2017. 

The numerical analysis was conducted  in \texttt{R}~\citep{R}.  We adapted the
\texttt{R}  package  \texttt{SuperLearner}~\citep{SuperLearner} in  a  package
called \texttt{SequentialSuperLearner}~\citep{SequentialSuperLearner}.

\subsection{Results}
\label{sec:appli:results}

\begin{figure}[htbp]
  \centering%
  \includegraphics[scale=1]{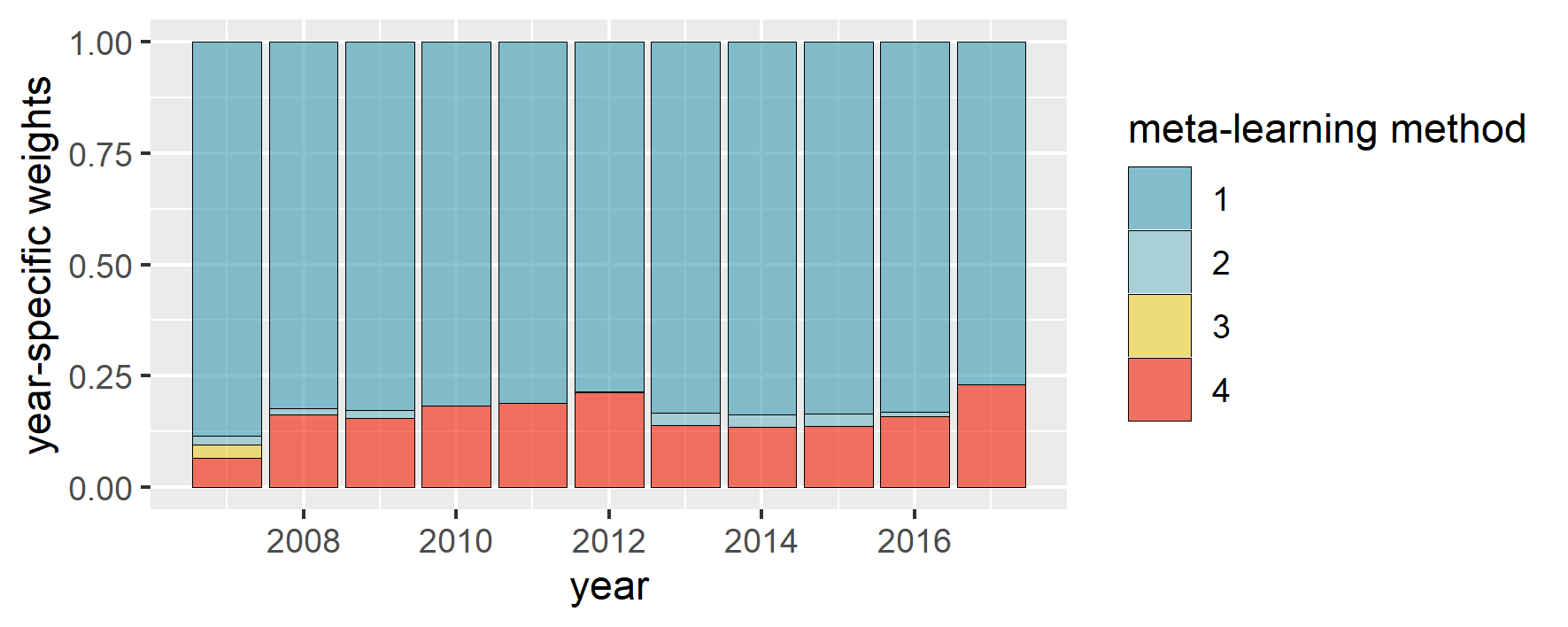}
  \caption{Evolution  (from 2007  onward)  of the  weights  attributed in  the
    overarching    Super     Learner    to    four    of     the    algorithms
    $\widehat{\theta}_{1},  \ldots \widehat{\theta}_{J}$.   The others  get no
    weight at all.}
  \label{fig:overarching-weights}
\end{figure}

In  Figure~\ref{fig:overarching-weights}  we  present  the  evolution  of  the
weights that characterize the continuous overarching Super Learner through the
years  2007  to  2017.  The  figure  reveals  that  only  four of  the  $J=50$
algorithms $\widehat{\theta}_{1}, \ldots  \widehat{\theta}_{J}$ get a positive
weight, and that only  two of them do in 2016 and 2017.   Moreover, one of the
algorithms dominates the  others during the whole training.  It  does not come
as  a surprise  that this  algorithm (whose  method is  a variant  of gradient
boosting  with  linear  boosters)  is  constantly  selected  by  the  discrete
overarching Super Learner.

For confidentiality  reasons, we were  not given the authorization  to discuss
how the  overarching Super Learners  fare compared to the  algorithm currently
deployed at CCR to predict the overall  costs of drought events in France from
2007 to 2017.  However,  we were authorized to make a  comparison for the sole
year  2017.  That  particular year,  the discrete  and continuous  overarching
Super  Learners outperform  the algorithm  currently deployed  at CCR,  with a
precision  of  96\% (discrete  overarching  Super  Learner), 94\%  (continuous
overarching Super Learners) versus 83\% (currently deployed algorithm).

\begin{figure}[htbp]
  \centering%
  \includegraphics[scale=0.65]{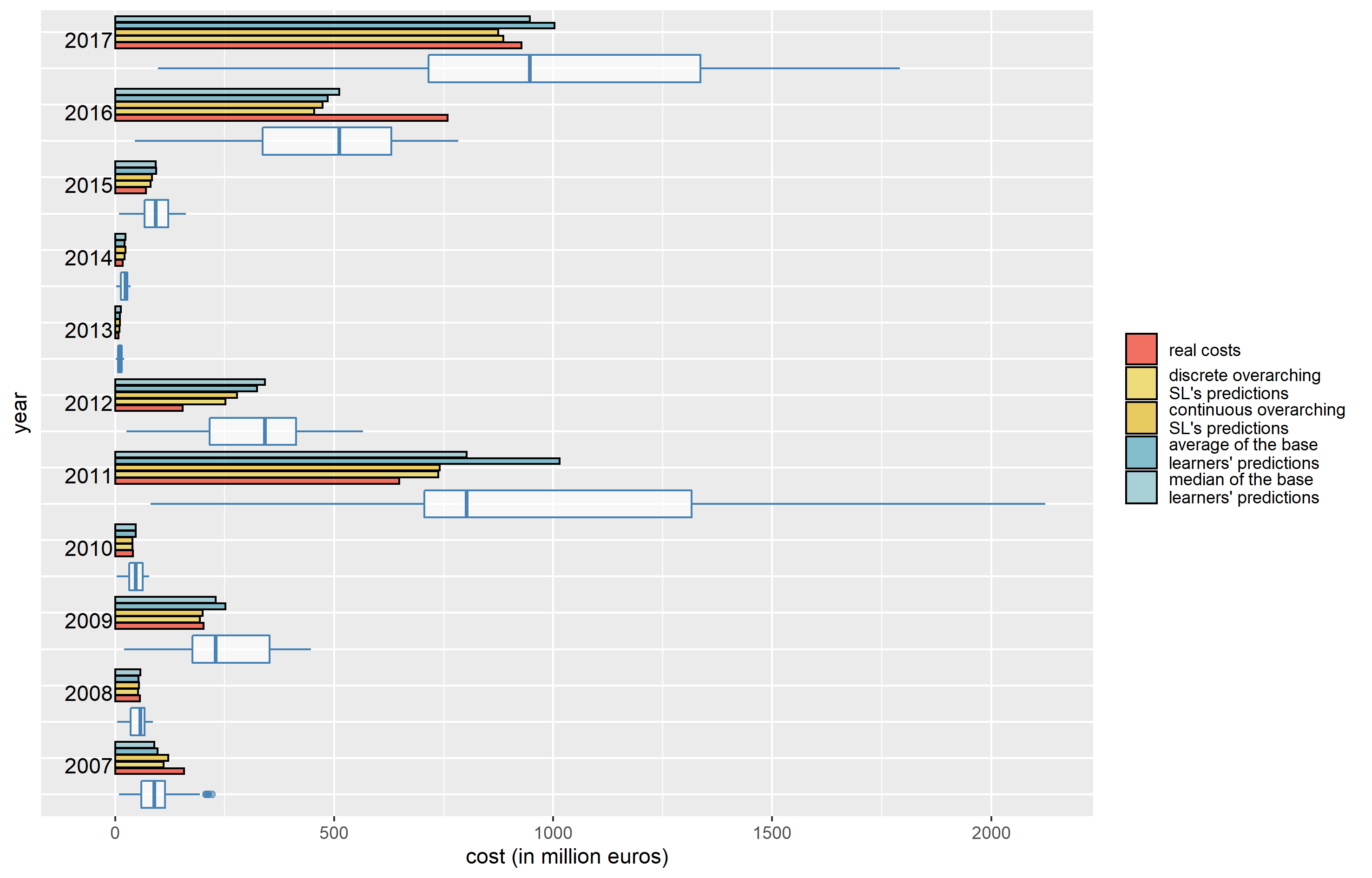}
  \caption{Presentation (from 2007 onward) of the real costs of drought events
    and  their predictions.   The predictions  are  either those  made by  the
    discrete  (pale yellow)  and  continuous overarching  (dark yellow)  Super
    Learners or obtained by averaging all the base learners' predictions (red)
    or  using their  median (blue  vertical  bars). The  figure also  presents
    boxplots that summarize all the base learners' predictions.  Note the high
    variability of these predictions. In this figure we use current euros.}
  \label{fig:predictions-overarching}
\end{figure}

In   Figure~\ref{fig:predictions-overarching}   we  primarily   present   four
sequences  of predictions  from  2007 to  2017: those  from  the discrete  and
continuous overarching Super Learners and those obtained by taking the average
or the  median of all  the base  learners' predictions.  Secondarily,  we also
summarize all the  base learners' predictions with  boxplots.  The variability
of  the base  learners'  predictions  is striking,  confirming  that the  base
learners can  strongly disagree.  Note  that the two sequences  of predictions
from  the Super  Learners are  quite  similar.  Overall,  the Super  Learners'
predictions look generally accurate and  better than the averaged predictions.
As for the medians of the predictions, they seem to provide a better trade-off
than the averages.  However neither the method consisting in using the average
of the  base learners' predictions  nor the  method consisting in  using their
median  is given  a  positive weight  by the  overarching  Super Learner.   In
Table~\ref{tab:mean:sd} we  report the averages and  standard deviations (over
the years)  of the ratios  of the  predicted costs to  the real costs  for the
predictors.   Both in  terms  of  mean and  standard  deviation, the  discrete
overarching Super Learner outperforms its continuous counterpart, which itself
outperforms the  predictors that average  or take the  median of all  the base
learners' predictions.   Furthermore, the two Super  Learners' predictions are
quite good for all years except 2012 and 2016.  The poorer predictions in 2016
are more  problematic because  the real cost  in 2016 is  much higher  than in
2012.

\begin{table}[h]
  \centering
  \begin{tabular}{r|cc}\hline
    predictions & mean & standard deviation\\\hline
    average of the base learners' predictions & 1.21 & 0.42\\
    median of the base learners' predictions & 1.21 & 0.45 \\
    continuous overarching Super Learner's & 1.10 & 0.32\\
    discrete overarching Super Learner's & 1.04 & 0.28\\\hline
  \end{tabular}
  \caption{Averages and standard deviations (over  the years) of the ratios of
    the predicted  costs to the real  costs. The predictions are  either those
    made by the discrete and continuous overarching Super Learners or obtained
    by  averaging  all the  base  learners'  predictions  or by  taking  their
    median. }
  \label{tab:mean:sd}
\end{table}

The  year  2016 is  known  in  the  French  insurance market  as  particularly
challenging. Unfortunately,  as far as we  know, this fact is  undocumented in
the literature.  However, we can report two facts to uphold this statement.


First, the year-specific  average cost is particularly large  in 2016 compared
to the global average cost between 2007 and 2017: 797,000 euros versus 482,000
euros. By year-specific  average cost we mean  the ratio of the  total cost of
the  year's   drought  event  to   the  corresponding  number   of  government
declarations of natural disaster for a  drought event delivered that year.  By
global average cost we mean the ratio  of the total cost of the drought events
between  2007 and  2017  to the  total number  of  government declarations  of
natural disaster for a drought event delivered these years.

Second, we  can quote~\citet[end of Section~4.1]{Charpentier_2022}  who say of
their predictions for the year 2016 that they are ``severely underestimated''.
Judging by their Figure~7, the  underestimation by the discrete and continuous
overarching  Super Learners  for the  year 2016  is less  pronounced than  the
underestimation by  their algorithms (but  we recall  that they tackle  a more
challenging problem  than us because we  focus on the city-specific  costs for
those cities that have obtained the government declaration of natural disaster
for a drought event whereas they consider all French cities).

\begin{figure}[htbp]
  \centering%
  \includegraphics[scale=0.8]{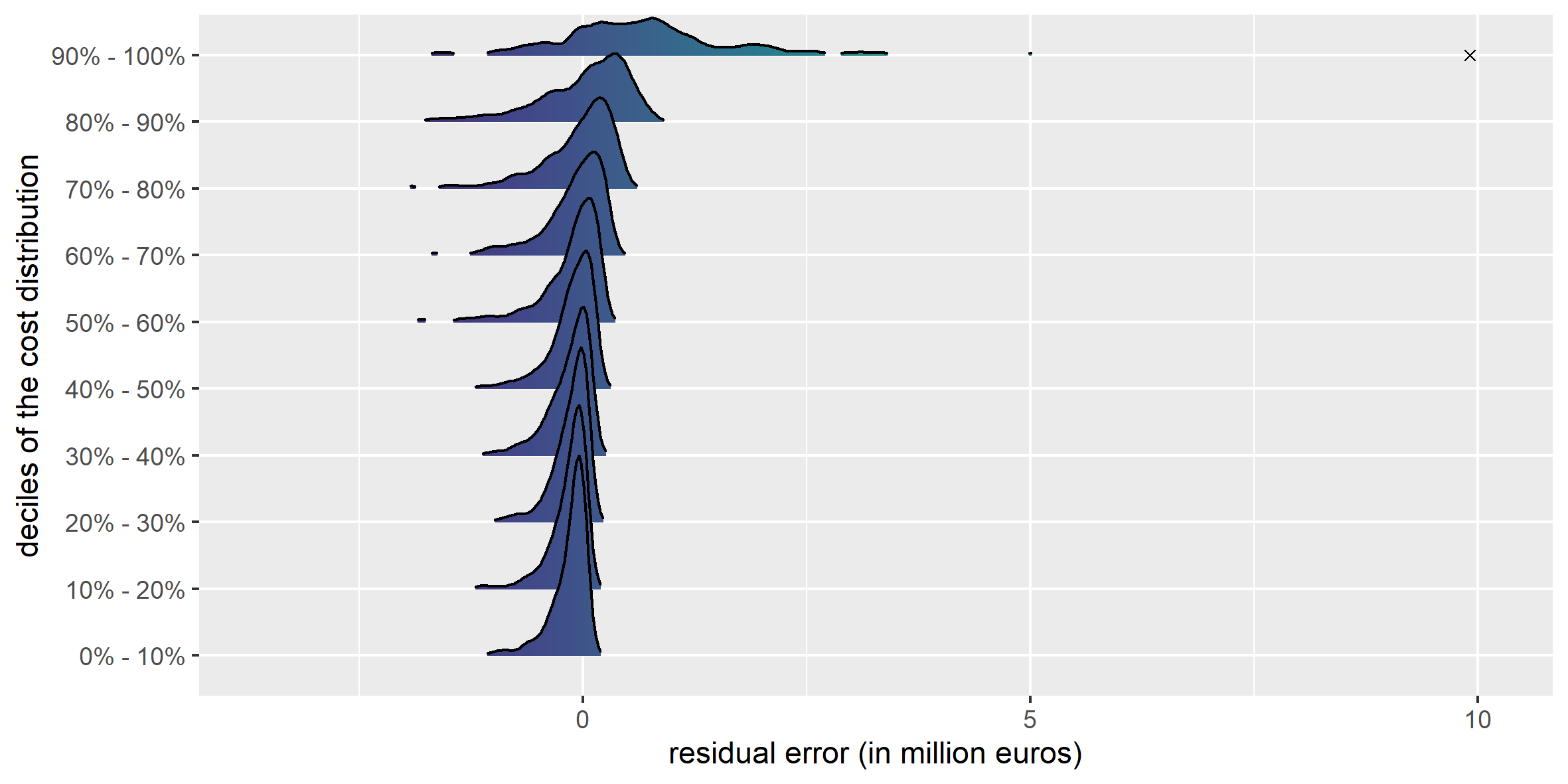}
  \caption{Kernel density  estimates of the  conditional laws of  the residual
    error  (of  the  predictions  made by  the  continuous  overarching  Super
    Learner)  in ten  strata characterized  by the  deciles of  the city-level
    costs. The  cross at the upper  right-hand side of the  plot indicates the
    maximum residual  error, made for a  city belonging to the  last decile of
    the cost distribution. In this figure we use current euros.}
  \label{fig:boxplot_errors}
\end{figure}

In  Figure~\ref{fig:boxplot_errors}  we   present  (Gaussian)  kernel  density
estimates of the  conditional laws of the residual error  (defined as the real
cost minus the prediction made by  the continuous overarching Super Learner --
the figure  is very similar  when substituting the discrete  overarching Super
Learner for the continuous one) in  ten strata characterized by the deciles of
the city-level costs. We note that the higher the city-level costs, the higher
the residuals.  Moreover, the overarching  Super Learner tends to overestimate
the costs in cities with lower city-level costs and, on the contrary, it tends
to underestimate them in cities with higher city-level costs.

\begin{figure}[htbp]
  \centering%
  \includegraphics[scale=0.65]{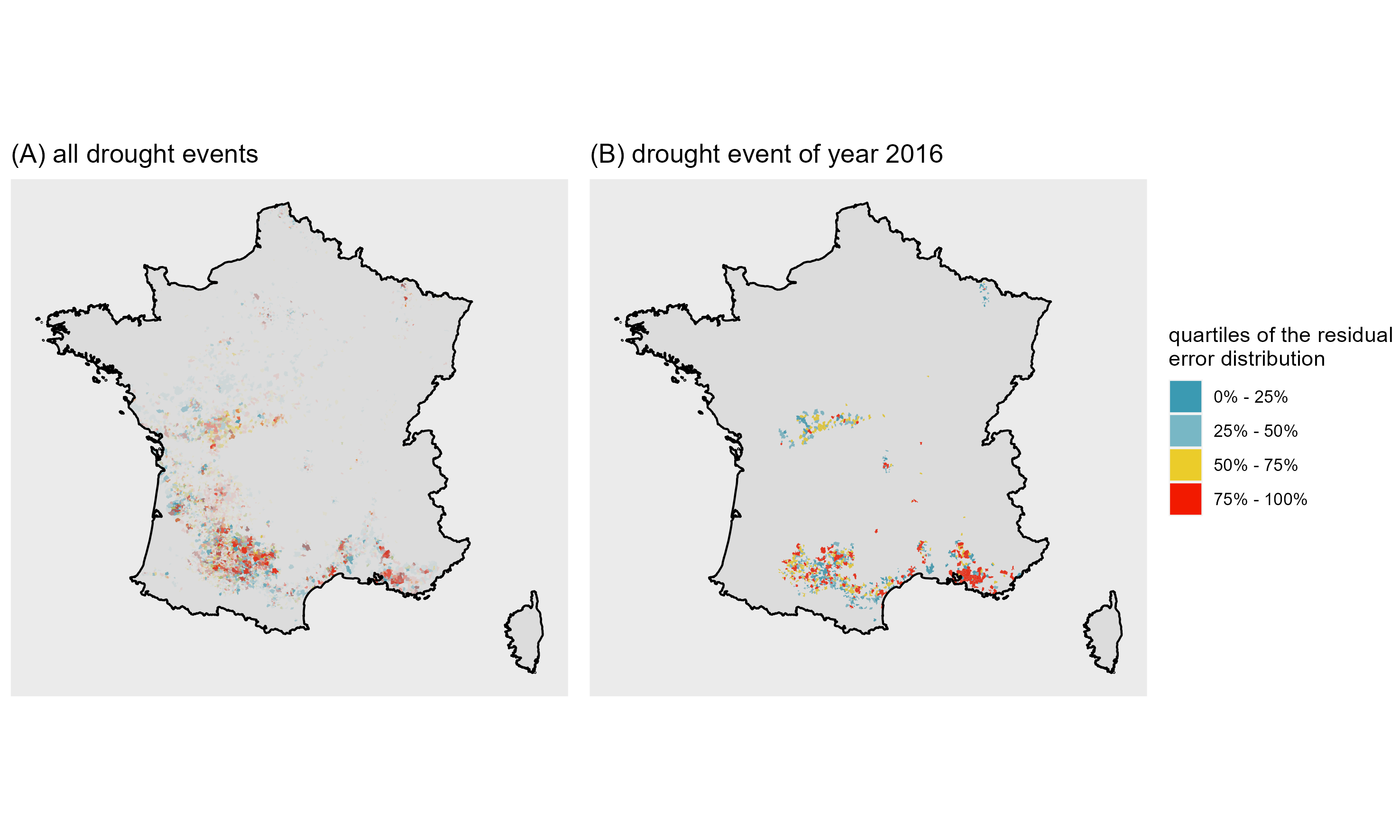}
  \caption{Geographical   distribution  of   the  residual   errors  (of   the
    predictions made by the continuous overarching Super Learner). (A): a city
    contributes as  many points  as the  number of times  it benefited  from a
    government declaration  of natural  disaster for  a drought  event between
    2007  and 2017.   (B):  a city  contributes  a  point if  and  only if  it
    benefited from a government declaration  of natural disaster for a drought
    event in 2016.   The color reflects the quartile of  the residual error to
    which the city- and time-specific residual error belongs (based on all the
    errors).  In (A),  the transparency reflects the number of  times the city
    benefited from a government declaration  of natural disaster for a drought
    event  between   2007  and   2017,  a  larger   number  leading   to  less
    transparency.}
  \label{fig:map_errors}
\end{figure}

In Figure~\ref{fig:map_errors} we  present two maps that  provide insight into
the geographical distribution of the  residual errors (of the predictions made
by the continuous overarching Super Learner  -- the maps are very similar when
considering  its discrete  counterpart). In  the  left-hand side  map, a  city
contributes  as  many points  as  the  number of  times  it  benefited from  a
government declaration  of natural disaster  for a drought event  between 2007
and 2017.  In the right-hand side map,  a city contributes a point if and only
if  it benefited  from  a government  declaration of  natural  disaster for  a
drought event  in 2016, the  year considered as particularly  challenging.  In
both maps, the color reflects the quartile  of the residual error to which the
city- and  time-specific residual error  belongs.  Moreover, in  the left-hand
side map the transparency reflects the number of times the city benefited from
a government declaration of natural disaster  for a drought event between 2007
and 2017, a larger number leading  to less transparency.  By comparing the two
maps,  we notice  \textit{(i)} that  the  2016 drought  episode impacted  very
strongly  the South  of France  and \textit{(ii)}  that, in  this region,  the
residual errors tend to be higher, leading to the underestimation of the local
cost.

\subsection{On the importance of the variables used to make predictions}
\label{subsec:var:imp}

The discrete and continuous overarching  Super Learners make predictions based
on a  multi-faceted description  of cities and  their exposures  consisting of
slightly  fewer  than  400  covariates (see  Section~\ref{sec:data}).   It  is
natural to wonder which variables mainly influence the prediction. 

The question pertains to the  definition and estimation of variable importance
measures.     The     literature    on     this    topic    is     rich    and
both~\citep{vdL2006,HKPvdL2016,williamson2021} on  the one hand  or \citep[and
references therein]{lundberg2017} on the other hand give insights about how to
answer it.  Unfortunately, building on these approaches is unrealistic because
our  data set  consists  of a  \textit{short}  time-series with  time-specific
observations  consisting of  \textit{many dependent}  data-structures and,  to
boot, because  we are interested in  a \textit{high} number of  covariates. We
thus propose the following simple approach tailored to our needs.

Recall that,  at each  time $t  \geq 1$, the  OSASSL outputs  the $t$-specific
estimator $\theta_{\widehat{j}_{t},t}$.  For  any $\alpha\in\calA$, evaluating
this  estimator at  $(X_{\alpha,t+1}, Z_{\alpha,t+1})$  yields the  prediction
$\hat{Y}_{\alpha,t+1}       :=      \theta_{\widehat{j}_{t},t}(X_{\alpha,t+1},
Z_{\alpha,t+1})$ of  the cost  $Y_{\alpha,t+1}$ of the  drought event  at time
$(t+1)$   for  city   $\alpha$.    Highlighting   that  $X_{\alpha,t+1}$   and
$Z_{\alpha,t+1}$   consist    of   (many)    covariates,   let    us   rewrite
$(X_{\alpha,t+1},Z_{\alpha,t+1}) =: (C_{s,\alpha,t+1} : 1\leq s \leq S+1)$. By
convention, $C_{S+1,\alpha,\tau}$ is the indicator  of whether or not the city
$\alpha$ obtained a  government declaration of natural disaster  for a drought
event on year  $\tau$.  Because we impose $\hat{Y}_{\alpha,\tau} =  0$ if that
is  the  case,  we  will  not   consider  the  importance  of  the  $(S+1)$-th
covariate.

Set  arbitrarily $1\leq  s \leq  S$ and  $t=2017$.  If  $s$ is  such that  the
covariate $C_{s,\alpha,\tau}$ can be treated as a continuous variable, then we
let  $\rho_{s}$  be  the  absolute  value of  the  Spearman  rank  correlation
coefficient~\cite[Section~8.5]{Hollander99}       computed      based       on
$((\hat{Y}_{\alpha,\tau}, C_{s,\alpha,\tau})  : \alpha\in\calA,  2007\leq \tau
\leq t)$.  If $s$ is such  that $C_{s,\alpha,\tau}$ takes $v$ values (in which
case $2 \leq v  \leq 5$), then we let $\rho_{s}$ be  the the correlation ratio
computed                                based                               on
$((\hat{Y}_{\alpha,\tau}, C_{s,\alpha,\tau})  : \alpha\in\calA,  2007\leq \tau
\leq t)$:
\begin{equation*}
  \rho_{s}:=\left(\frac{\sum_{\nu=1}^{v}     n_{\nu}      (\bar{y}_{\nu}     -
      \bar{y})^{2}}{\sum_{\alpha\in\calA}\sum_{\tau=2007}^{t}
      (\hat{Y}_{\alpha,\tau} - \bar{y})^{2}}\right)^{1/2}
\end{equation*}
where $\bar{y}_{\nu}$ is the average of the $\hat{Y}_{\alpha,\tau}$s such that
$C_{s,\alpha,\tau}   =   \nu$   and   $\bar{y}$  is   the   average   of   all
$\hat{Y}_{\alpha,\tau}$s.   Note  that we  could  have  defined $\rho_{s}$  as
Wilcoxon test's statistic (case $v=2$) or the Kruskal-Wallis test's statistics
(case $3\leq v\leq 5$)~\cite[see][Sections~3.1 and 6.1]{Hollander99} but chose
not  to, preferring  that all  $\rho_{s}$s naturally  lie in  $[0,1]$ to  ease
comparisons.

In all cases the larger is $\rho_{s}$  the more we are willing to believe that
the $s$-th covariate  well explains the predictions made by  the OSASSL.  Note
how we substituted the word ``explains''  for the word ``influences''. This is
an acknowledgement that  our assessments simply rely on  associations and have
no  causal  interpretation.    We  resort  to  permutation   tests  to  assess
significance levels, with one million independent permutations drawn uniformly
in each of  the above cases.  For  every $1\leq s \leq S$,  $\rho_{s}$ is much
larger (often by several orders of  magnitude) than the maximum value obtained
by permutation.  This strongly supports the findings reported below. 
        
\begin{figure}
  \centering
  \includegraphics*[scale=0.8]{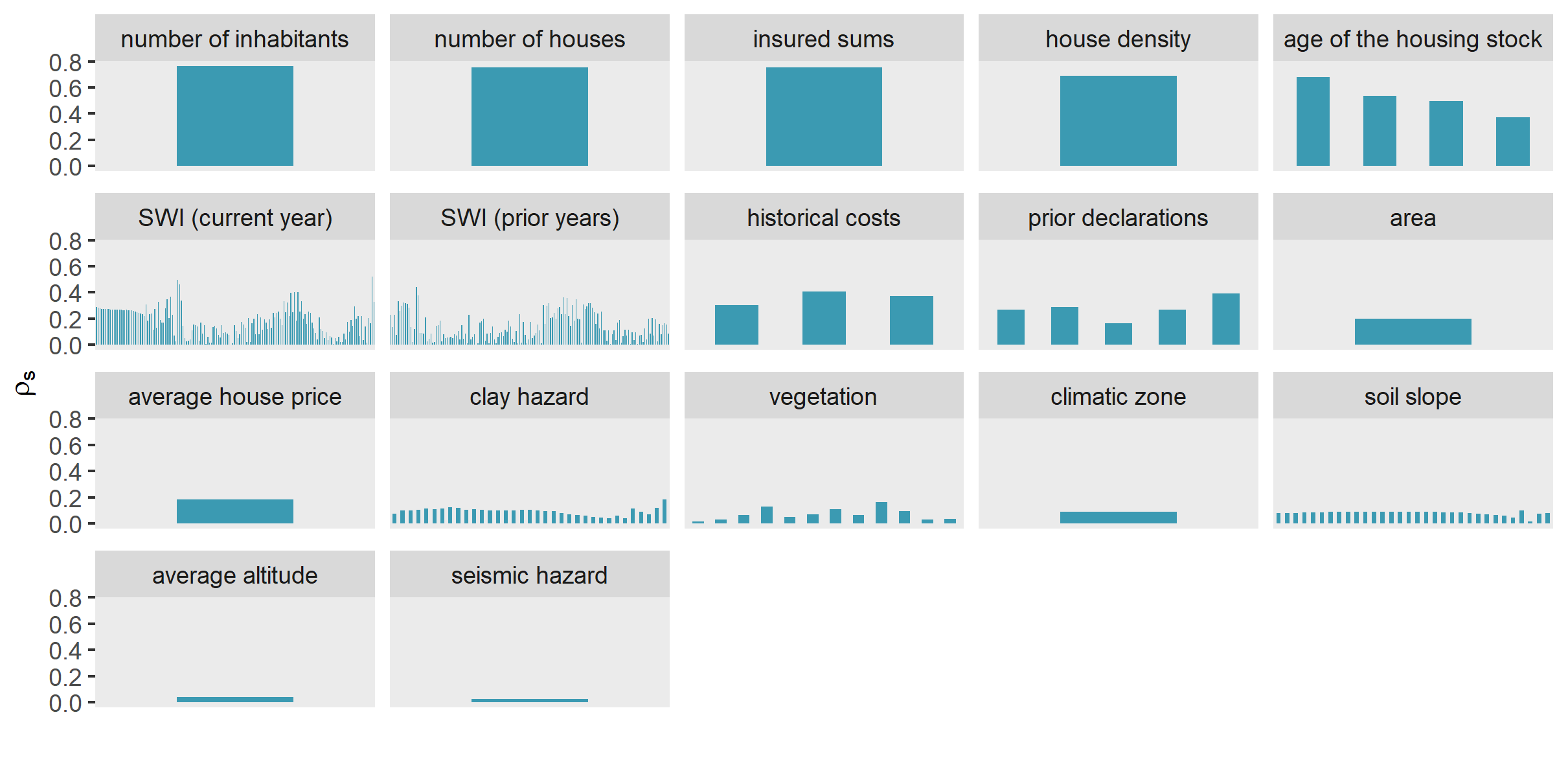}
  \caption{Assessing the importance of the  variables used to make predictions
    by the  discrete overarching Super  Learner. The larger is  $\rho_{s}$ the
    more we are willing to believe that the $s$-th covariate well explains the
    predictions  made  by  the  discrete  overarching  Super  Learner.   Every
    $\rho_{s}$  is  declared  significantly  positive by  a  permutation  test
    analysis.}
  \label{fig:var:imp}
\end{figure}

Surprisingly,  all  the   covariates  are  deemed  important   (based  on  the
permutation tests), though  some covariates are of course  more important than
others.  We  only report the  values of $(\rho_{s} :  1\leq s\leq S)$  for the
discrete overarching Super Learner (those  of the continuous overarching Super
Learner  are very  similar).  To  do  so, we  regroup the  covariates into  17
homogeneous categories: nine of them consist of a single covariate (the city's
area, average  altitude, average  house price,  climatic zone,  house density,
insured sums, number  of inhabitants, number of houses,  seismic hazard); four
of them consist of  a small number of covariates grouped  by theme (related to
the  city's age  of  the  housing stock,  historical  costs, prior  government
declarations of  natural disaster  for a drought  event, vegetation);  four of
them gather many covariates by theme  (related to the city's clay hazard, soil
slope,    SWI   during    the   current    year   and    history   of    SWI).
Figure~\ref{fig:var:imp} summarizes the results.  The most important variables
(number  of inhabitants,  number of  houses, insured  sums) are  related to  a
potential of exposure  to drought events.  The next two  more important (group
of)  variables  (house  density  and  age   of  the  housing  stock)  help  to
characterize the buildings at risk. The  variables that we mentioned so far do
not vary  significantly over  time.  The  next two  more important  (group of)
variables (SWI, current and prior  years) provide a meteorological description
of  the drought  events,  which  obviously vary  over  time.  As  anticipated,
relying on past descriptions of the drought events is relevant.  The remaining
(groups of) variables complete the characterization of the buildings at risk.

\conclusions[Discussion]  
\label{sec:discussion}

The French legal framework known  as the natural disasters compensation scheme
was created  in 1982. Drought  events were included  in 1989. Since  then they
have  been the  second most  expensive type  of natural  disaster.  In  recent
years, drought events have been remarkable in their extent and intensity.  The
problem  is worsening  and not  limited  to France,  as was  predicted in  the
technical  report~\citep[page~7]{SwissRe2011}: ``as  our climate  continues to
change, the  risk of property  damage from  soil subsidence [that  is, drought
events]  is not  only  increasing but  also spreading  to  new regions  across
Europe''.

Forecasting the cost of a drought event is an important actuarial problem.  To
tackle this  challenge, we develop  a new  methodology that builds  upon Super
Learning,  a  popular aggregation  strategy.   Our  overarching Super  Learner
blends predictions  made by a  collection of OSASSLs which,  themselves, blend
the predictions made by a variety of machine-learning algorithms.

We   introduced   and   studied   the   theoretical   properties   of   OSASSL
in~\citep{osasl}.   The   theoretical  analysis   hinges  on   a  stationarity
assumption stating that the  mechanism producing a local drought-event-related
cost conditionally on  its local description remains  constant throughout time
and  France.  The  assumption  warrants  both the  possibility  to define  and
estimate  the mean  conditional  cost on  the  one  hand and  the  use of  its
estimator to  make predictions on the  other hand. Predictions can  be made at
any local  description $(x,z)$  provided that  it falls in  the domain  of the
observed local descriptions.  Naturally, the scarcer the available information
around $(x,z)$,  the less  reliable the prediction.   In addition,  if $(x,z)$
falls outside the domain then, although a prediction may be made nevertheless,
it cannot be trusted.  Therefore, in  view of climate change, it is meaningful
to make  projections of drought  events in the not-too-distant  future.  Under
another assumption on the complex dependence  structure induced in the data by
the spatial and  temporal nature of the phenomenon of  drought, we showed that
OSASSL can learn the mean conditional cost, making up for the shortness of the
time-series thanks to  the manyness of each  time-specific observation because
the latter are only slightly dependent.

In  this article,  we focus  on  the application  of OSASSL.   We present  two
implementations,  called   the  discrete  and  continuous   overarching  Super
Learners.   Their predictions  are generally  accurate and  better than  those
obtained, for instance, by averaging or taking the median of all the low-level
predictions made  by the base  machine-learning algorithms (two ways  among 50
implemented to combine  the 27 low-level predictions).   Specifically, the two
Super Learners' predictions are quite good for all years except 2012 and 2016.
The poorer predictions in 2016, a year known in the French insurance market to
be particularly  challenging, are  more problematic because  the real  cost in
2016 is much  higher than in 2012.  Moreover, we  were given the authorization
to compare  the predictions of  the discrete and continuous  overarching Super
Learners with  that of the  algorithm currently deployed  at CCR for  the sole
year 2017:  the precisions are  respectively 96\% (discrete  overarching Super
Learner),  94\% (continuous  overarching Super  Learners) and  83\% (currently
deployed algorithm).

The quality of the predictions made by the overarching Super Learners strongly
depends  on  the  relevance  and  quality  of  the  covariates  used  to  make
predictions --- in particular, on the  local description of the drought event.
Regarding the covariates' relevance, we develop an \textit{ad hoc} approach to
define  and  estimate variable  importance  measures  so  as to  assess  which
covariates mainly  influence the  predictions.  We  acknowledge that  the word
``influence'' is somewhat misleading because a causal interpretation is out of
reach and our  assessments simply rely on associations.   Surprisingly, all of
the covariates are  deemed relevant (based on permutation  tests), though some
covariates are more important than others.  Regarding the covariates' quality,
the overarching Super  Learners would probably benefit from  a refined version
of the city-level  SWI that, contrary to  the one we rely on,  does not assume
that the  nature of the soil  is the same  all over France.  In  addition, the
local  description  would  also  be   considerably  enhanced  if  it  included
information such  as the distribution of  the proximity between a  house and a
tree at the city-level, or the  distribution of the depth of house foundations
at the city-level.   Such pieces of information are proxies  to soil shrinkage
and  swelling.  The  local description  could  also be  enhanced by  including
direct measurements  of soil shrinkage and  swelling which can be  obtained by
radar interferometry.

In this  study, we  forecast the  cost of  drought events  in France  by Super
Learning for  those cities  that have obtained  the government  declaration of
natural disaster for a drought event.  The  next step will be to predict which
cities  will obtain  the  government  declaration of  natural  disaster for  a
drought event.  Tackling  this difficult challenge will  allow forecasting the
cost of drought events earlier.

\codeavailability{The     numerical     analysis      was     conducted     in
  \texttt{R}~\citep{R}.      We     adapted     the     \texttt{R}     package
  \texttt{SuperLearner}~\citep{SuperLearner}     in    a     package    called
  \texttt{SequentialSuperLearner}~\citep{SequentialSuperLearner}.}

\noappendix

\appendixfigures

\appendixtables 

\authorcontribution{The authors contributed equally to this work.} 

\competinginterests{The authors contributed equally to this work.}

\begin{acknowledgements}
  The authors thank Thierry Cohignac (Caisse Centrale de Réassurance) and Herb
  Susmann (MAP5 and Department of Biostatistics \& Epidemiology, University of
  Massachusetts Amherst) for their suggestions.
\end{acknowledgements}

\bibliographystyle{copernicus}
\bibliography{super_learning}

\end{document}